\documentclass[journal]{IEEEtran}

\ifCLASSINFOpdf

\else

\fi

\usepackage{graphicx}
\usepackage{amsthm,amssymb,amsfonts}
\usepackage{epsfig} % for postscript graphics files
\usepackage{epstopdf}
\usepackage{amsmath}
\usepackage{amsthm}
\usepackage{comment}
\usepackage{url}
\usepackage{algorithm}
\usepackage{algorithmic}
\usepackage{subfigure}
\usepackage{booktabs}

\usepackage{amsfonts}
\usepackage{bm}
\usepackage{color}
\usepackage{xcolor}
\usepackage{cite}
\usepackage{stfloats}
\usepackage{chemarrow}
\usepackage{extarrows}
\usepackage{setspace}

\usepackage{microtype}
\usepackage[american]{babel}

\usepackage[explicit]{titlesec}
\allowdisplaybreaks

\newboolean{showcomments}
\setboolean{showcomments}{true}
\newcommand{\wang}[1]{\ifthenelse{\boolean{showcomments}}
	{ \textcolor[rgb]{1,0,1}{(ZW:  #1)}}{}}
\newcommand{\fliu}[1]{\ifthenelse{\boolean{showcomments}}
	{ \textcolor{blue}{(FL:  #1)}}{}}
\newcommand{\ychen}[1]{\ifthenelse{\boolean{showcomments}}
	{ \textcolor{green}{(ZP:  #1)}}{}}
\newcommand{\slow}[1]{\ifthenelse{\boolean{showcomments}}
	{ \textcolor{blue}{(SL:  #1)}}{}}

\theoremstyle{definition}

\theoremstyle{definition}

\def\BibTeX{{\rm B\kern-.05em{\sc i\kern-.025em b}\kern-.08em
		T\kern-.1667em\lower.7ex\hbox{E}\kern-.125emX}}

\title{Incentive Mechanism Design for Emergency Frequency Control in Multi-Infeed Hybrid \\AC-DC System}

\begin{document}
\setstretch{1}
\author{
	Ye~Liu,~\IEEEmembership{Student Member, IEEE}, 
	Chen~Shen,~\IEEEmembership{Senior Member, IEEE}, 
	Zhaojian~Wang,~\IEEEmembership{Member, IEEE}, 
	\\and Feng~Liu,~\IEEEmembership{Senior Member, IEEE}  
        %Steven~Low,~\IEEEmembership{Fellow,~IEEE} and Shengwei~Mei,~\IEEEmembership{Fellow,~IEEE}
        % <-this % stops a space
        \thanks{(Corresponding author: Chen Shen.)}     

		\thanks{Ye Liu, Chen Shen, and Feng Liu are with the State Key Laboratory of Power Systems, Department of Electrical Engineering, Tsinghua University, Beijing 100084, China (e-mail: liuye18@mails.tsinghua.edu.cn; shenchen@mail.tsinghua.edu.cn; lfeng@mail.tsinghua.edu.cn).}
		
		\thanks{Zhaojian Wang is with the Key Laboratory of System Control, and Information Processing, Ministry of Education of China, Department of Automation, Shanghai Jiao Tong University, Shanghai 200240, China (e-mail: wangzhaojian@sjtu.edu.cn).}
			
			% <-this % stops a space
}

%% The paper headers
\markboth{Journal of \LaTeX\ Class Files,~Vol.~xx, No.~xx, xx~xxxx}%
{Shell \MakeLowercase{\textit{et al.}}: Bare Demo of IEEEtran.cls for IEEE Journals}

\maketitle
%\thispagestyle{empty}
%\pagestyle{empty}

\begin{abstract}
    In multi-infeed hybrid AC-DC (MIDC) systems, the emergency frequency control (EFC) with LCC-HVDC systems participating is of vital importance for system frequency stability. Nevertheless, when regional power systems are operated by different decision-makers, the LCC-HVDC systems and their connected AC systems might be unwilling to participate in the EFC due to the costs and losses. In this paper, to incentivize the LCC-HVDC systems and their connected adjacent AC systems to participate in the droop-based EFC, a novel control-parameter-based incentive mechanism is proposed, which can deal with various possible emergency frequency faults. Then, a non-cooperative-based incentive game model is formulated to implement the incentive mechanism in the MIDC system. An algorithm for seeking the Nash equilibrium is designed, and the uniqueness of Nash equilibrium is proven. Moreover, the individual rationality, incentive compatibility and social optimality of the proposed mechanism are analyzed and proven. The effectiveness of the proposed incentive mechanism is verified through a case study.  
\end{abstract}


%\IEEEpeerreviewmaketitle

\section{Introduction}

\subsection{Motivation}

Conventional AC power systems have been gradually transformed into complex hybrid AC-DC power systems due to the development of HVDC technologies \cite{wang2013harmonizing,litzenberger2016s}. In China, the line-commutated-converter-based HVDC (LCC-HVDC) systems \cite{kimbark1971direct} are widely applied as the tie-lines between regional grids, which forms asynchronous multi-infeed hybrid AC-DC (MIDC) systems \cite{zhou2018principle}. In an MIDC system, one AC main system is connected with multiple adjacent AC systems through multiple LCC-HVDC systems (refer to Fig. \ref{midc_topo} for detailed topology). The complex dynamics of the MIDC system bring challenges to the stable operation.

Considering the frequency stability of the MIDC system, the traditional frequency regulation strategies might not meet the frequency stability requirements because: 1) the feeding of multiple LCC-HVDC systems might lead to the frequency regulation reserve shortage and inertia shortage of the AC main system, 2) emergency faults (e.g., HVDC block faults or AC-DC cascading faults) with considerable power imbalances are prone to occur in the MIDC system \cite{kwon2020optimal,bevrani2009robust}. Therefore, the MIDC system requires emergency frequency control (EFC) strategy. Compared with the conventional EFC strategies, i.e., the load shedding and generator tripping operations \cite{terzija2006adaptive,song2016review}, the EFC strategies with LCC-HVDC systems participating are more economic and effective for the MIDC system, which utilize the active power fast adjustability \cite{harnefors2016impact} and the short-time overload capacity \cite{aidong2007study} of the LCC-HVDC system. 

Considering the EFC strategies in MIDC systems, the existing works focus on how to design EFC strategies to meet various control objectives in engineering practice \cite{liu2017design,du2012integrated}, under the assumption that the entire power system is operated by one decision-maker and all the LCC-HVDC systems and adjacent AC systems are obliged to participate in the EFC of the AC main system. For instance, in our previous work \cite{liu2021optimal}, a decentralized droop-based EFC strategy is proposed. Nevertheless, considering the future development and marketization of the power grid \cite{litvinov2019electricity,singh2003art}, the AC main system, LCC-HVDC systems and adjacent AC systems might belong to different decision-makers. In addition, participating in the EFC brings generation costs, reserve costs and frequency deviation losses to the adjacent AC systems, and also brings equipment losses caused by fast power regulation to the LCC-HVDC systems. Therefore, the adjacent AC systems and LCC-HVDC systems might be unwilling to participate in the EFC due to their self-interests, which need incentives from the AC main system. In this paper, to reasonably incentivize the adjacent AC systems and LCC-HVDC systems to participate in the droop-based EFC, a novel control-parameter-based incentive mechanism is proposed. Note that the design idea of the proposed incentive mechanism can be extended to other EFC strategies, and the incentive mechanism can also provide reference for the power grid with single decision-maker.     

\subsection{Literature Review}

To the best of our knowledge, there is no relevant research about the incentive mechanism design for the EFC with LCC-HVDC systems participating, but there are various design ideas of incentive mechanism for frequency control in the existing literature. In \cite{liu2012performance}, a control performance standard (CPS)-based incentive mechanism is proposed to incentivize power producers to participate in the automatic generation control (AGC). However, the CPS index is defined based on the secondary frequency control in synchronous AC power systems, thus, the mechanism in \cite{liu2012performance} is not applicable to the EFC of asynchronous MIDC systems. In \cite{zhu2014stability}, an incentive mechanism for distributed frequency control implementation is proposed to enforce the system stability, which is based on the Vickrey-Clarke-Groves (VCG) mechanism \cite{makowski1987vickrey}. Nevertheless, the VCG-based mechanism is not self-budget balancing and extra bonuses are required to balance the budget. 

Subsequently, we introduce the relevant works about the frequency regulation reserve market mechanism, since which can be regarded as a reserve-capacity-based incentive mechanism for frequency control \cite{wang2019incentive}. There are two significant steps in the frequency regulation reserve market mechanism, which are reserve capacity determination and reserve market clearing, respectively \cite{borne2018barriers}. The methods for reserve capacity determination mainly include the reliability-based method \cite{najafi2010optimal}, the conditional-value-at-risk (CVaR) based approach \cite{chen2014optimal} and the method based on cost-benefit analysis \cite{zhang2000study,afshar2008cost}, etc. And the reserve market clearing methods mainly contain the bilateral-contract-based method \cite{madrigal2000security}, the game-based approach \cite{haghighat2007gaming} and the bidding-based approach \cite{campos2015strategic}, etc. The above various methods make the frequency regulation reserve market mechanism widely applied in the engineering practice, but this kind of reserve-capacity-based incentive mechanism must be cleared according to a specific reserve capacity demand. Considering the unpredictable emergency frequency faults with random occurrences and considerable power imbalances, the aforementioned reserve-capacity-based approach usually conservatively selects the severest faults as the reserve capacity demand for market clearing, which causes economic losses and waste of reserve resources. Therefore, the less conservative incentive mechanisms which can deal with various emergency faults are urgently required.

Moreover, considering the incentive mechanisms for the demand side with frequency regulation potential, in \cite{wang2020reserve}, a user-oriented double-incentive mechanism is proposed to improve the reserve service capability of electric vehicles (EVs), in which the semi-managed response mode effectively simplifies the response process. In \cite{liu2014full,konda2017optimal}, the co-optimization method for demand response aggregators to participate in the energy and reserve market is studied. The above works focus on the optimal decision-making and operation methods for the demand side resources when involved in the frequency regulation market, so as to incentivize them to participate in the frequency regulation. Nevertheless, the system frequency security constraints are barely considered in the above research, which are supposed to be considered in the incentive mechanism for EFC.    

\subsection{Contribution}

According to the literature review, to design a incentive mechanism which is applicable to the droop-based EFC of asynchronous MIDC systems, the key challenge is how to enable the incentive mechanism to deal with various unpredictable emergency frequency faults, with frequency security constraints considered. In this paper, a control-parameter-based idea is utilized to design the incentive mechanism for droop-based EFC, and a non-cooperative game model (i.e., the incentive game) is formulated to implement the proposed mechanism. Benefitting from the control parameter immediate adjustability of the droop-based EFC, the proposed mechanism can handle various possible emergency faults. In addition, this incentive mechanism utilizes the  short-time overload capability of LCC-HVDC systems, which ensures the operation economy of LCC-HVDC systems in normal conditions. The contributions of this paper are summarized as follows:
\begin{itemize}
	\item A novel control-parameter-based incentive mechanism is proposed to incentivize the adjacent AC systems and LCC-HVDC systems to participate in the droop-based EFC strategy. Different from the reserve-capacity-based approach in \cite{madrigal2000security,haghighat2007gaming,campos2015strategic}, the proposed mechanism can deal with various emergency frequency faults.
	%In this mechanism, each adjacent AC system obtains a part of a fixed reward provided by the AC main system, and each part of the reward is proportional to the droop coefficient of the connected LCC-HVDC system. Benefitting from the immediate adjustability of the droop coefficient, the EFC strategy under this mechanism can handle various possible emergency faults. The utilization of the LCC-HVDC short-time overload capacity ensures the operation economy of the LCC-HVDC system in normal conditions.   
	\item To implement the proposed incentive mechanism, a non-cooperative-based incentive game model is proposed and formulated, with frequency security constraints considered. Then, an algorithm for seeking the Nash equilibrium of the incentive game is designed for practical engineering applications, and the uniqueness of the equilibrium is proven.
	\item Considering the properties of the proposed incentive mechanism, the individual rationality and the incentive compatibility are analyzed and proven. Moreover, the Nash equilibrium of the incentive game is rigorously proven to be the social optimum of the defined social welfare problem.
\end{itemize}

\subsection{Organization}
The rest of this paper is organized as follows. Section II introduces the control-parameter-based incentive mechanism for droop-based EFC strategy. Section III formulates and analyzes the incentive game model. Section IV discuss the properties of the proposed incentive mechanism. In Section V, an MIDC system case is tested and the effectiveness of the proposed incentive mechanism is verified. Section VI concludes this paper.

%\textcolor{red}{.}
\section{Incentive Mechanism for Droop-Based Emergency Frequency Control}
In this section, we first introduce some preliminaries on the droop-based EFC strategy in the MIDC system. Then, the control-parameter-based incentive mechanism for droop-based EFC strategy is proposed.

\begin{figure}[tb]
	\centering
	%\vspace{-0.2cm}  %调整图片与上文的垂直距离
	%\setlength{\abovecaptionskip}{-0.05cm}   %调整图片标题与图距离
	%\setlength{\belowcaptionskip}{-2cm}   %调整图片标题与下文距离
	\includegraphics[width=0.45\textwidth]{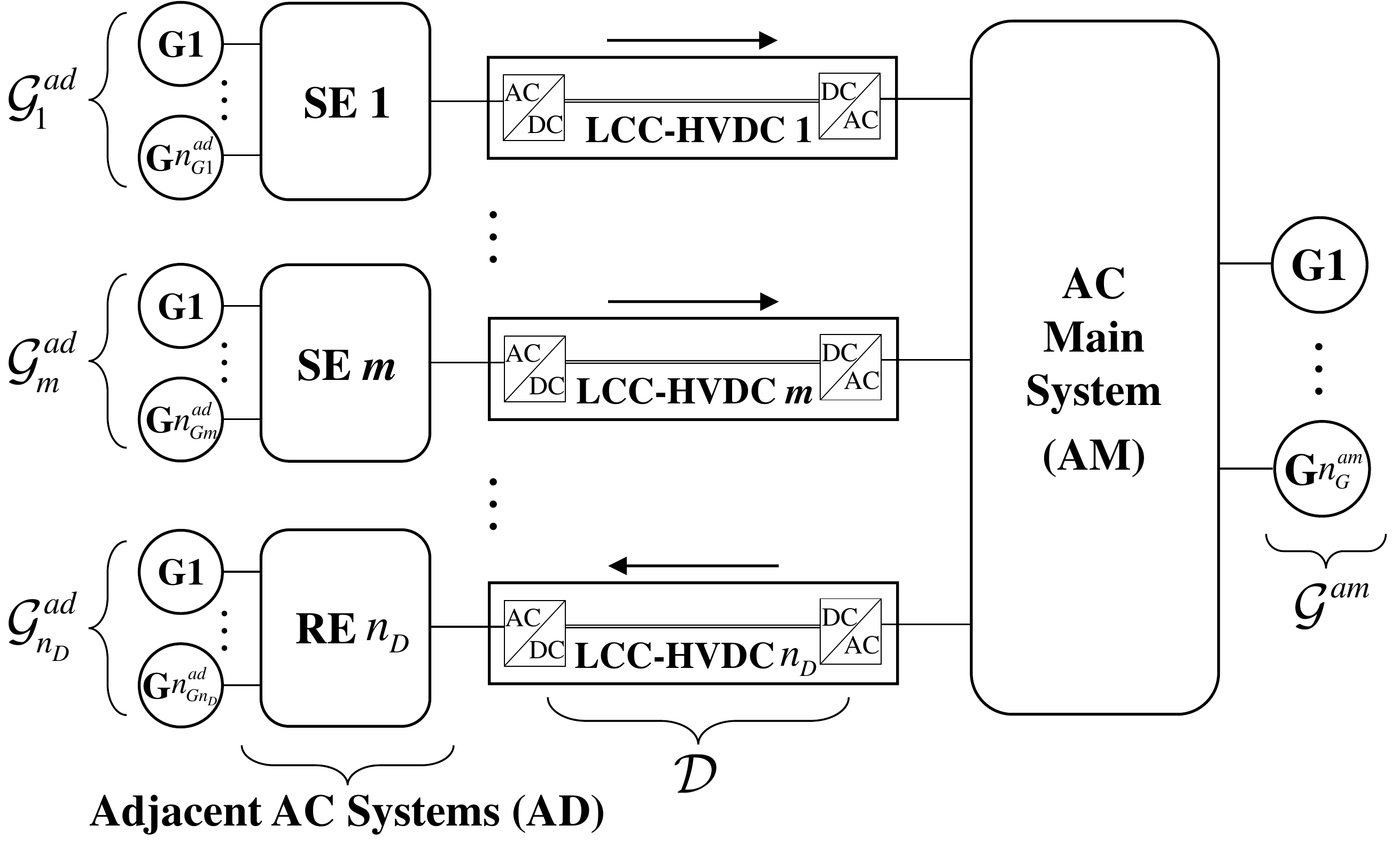}
	\caption{Topology of the MIDC system.}
	\label{midc_topo}
\end{figure}

Generally, the topology of an asynchronous-interconnected MIDC system is shown in Fig. \ref{midc_topo}, where the AC main system (AM system for short) is connected with $n_D$ LCC-HVDC systems. Specifically, there are $m$ LCC-HVDC systems transmitting power from sending-end (SE) systems to the AM system and ($n_D - m$) LCC-HVDC systems transmitting power from the AM system to receiving-end (RE) systems, which are called SE-LCC systems and RE-LCC systems respectively. The SE 1$\sim$$m$ and RE ($m$+1)$\sim$$n_D$ systems are collectively called the adjacent AC systems (AD systems for short). For the topology of the MIDC system, we make the following assumptions:

\textbf{Assumption 1:} 1) One AD system is only connected with one LCC-HVDC system. In engineering practice, there exist multiple LCC-HVDC systems connected to the same AD system, and we can make them equivalent to one LCC-HVDC system. 2) The AD systems operate asynchronously. 

The set of LCC-HVDC systems is denoted by $\mathcal{D}$, which can also represents the set of AD systems due to Assumption 1. The set of generators in the AM system is denoted by $\mathcal{G}^{am}$, and the set of generators in AD $i$ system is denoted by $\mathcal{G}^{ad}_i$, where $i \in \mathcal{D}$. We denote the numbers of generators in the above sets with $n_G^{am}$ and $n_{Gi}^{ad}$, respectively. Based on the described MIDC system, we design the incentive mechanism for droop-based EFC strategy.

\subsection{Preliminaries on Droop-Based EFC Strategy}

The authors have proposed a decentralized droop-based EFC strategy to address the emergency frequency problems in the MIDC system \cite{liu2021optimal}. In \cite{liu2021optimal}, an active power-frequency (P-f) droop control strategy is designed and implemented for the LCC-HVDC system. Then, the droop-based EFC strategy is introduced, and the coordinated droop mechanism ensures that the LCC-HVDC droop control only works in emergency situations as supplementary support for conventional generators' primary frequency control. Moreover, considering the designed EFC strategy, a Lyapunov-function-based stability analysis illustrates the asymptotic stability of the closed-loop equilibrium. Benefitting from the decentralized design logic, the droop-based EFC strategy guarantees a fast response in case of emergency faults, and can easily be applied to practical engineering projects. The P-f droop control equation of LCC-HVDC $i$ is shown as follows:
\begin{align}
\label{lcc_droop_equ}
P_{ord,i}^{D} = P_{i}^D - k_i^D\omega^{am},\ i \in \mathcal{D}
\end{align} 
where $P_{i}^D$ is the nominal active power of LCC-HVDC $i$, which is $>0$ for SE-LCC or $<0$ for RE-LCC, $P_{ord,i}^{D}$ is the active power control order of LCC-HVDC $i$, $k_i^D>0$ is the droop coefficient of LCC-HVDC $i$, and $\omega^{am}$ is the frequency deviation from the nominal frequency of the AM system. 

In the scenario where the AM system and AD systems are operated by different decision-makers, supposing that the AM system has requirements for the steady-state frequency after emergency faults, i.e., $\omega^{am}$ is a given constant, and ignoring the upper and lower bounds of $P_{ord,i}^{D}$, the power support of LCC-HVDC $i$ under the designed EFC strategy is $\Delta P_{i}^D = -k_i^D\omega^{am}$, which is proportional to $k_i^D$. Thus, the contribution of LCC-HVDC $i$ to the frequency stability of the AM system can be characterized by its droop coefficient $k_i^D$. Then, we design the control-parameter-based incentive mechanism in the following subsection, and further explain that the design idea can be extended to various EFC strategies. 

\subsection{Control-Parameter-Based Incentive Mechanism Design}

In the design of incentive mechanism, we first make the following assumption:

\textbf{Assumption 2:} 1) Considering the consistency between AD $i$ system and its connected LCC-HVDC $i$ in decision-making and utility, we regard them as one decision-maker, and represent this decision-maker as AD $i$ in the mechanism design. 2) The LCC-HVDC systems and generators can provide enough power support to cover the power imbalances caused by emergency faults, and so the load shedding operations are ignored in the mechanism design.

\begin{figure}[tb]
	\centering
	%\vspace{-0.2cm}  %调整图片与上文的垂直距离
	%\setlength{\abovecaptionskip}{-0.05cm}   %调整图片标题与图距离
	%\setlength{\belowcaptionskip}{-2cm}   %调整图片标题与下文距离
	\includegraphics[width=0.44\textwidth]{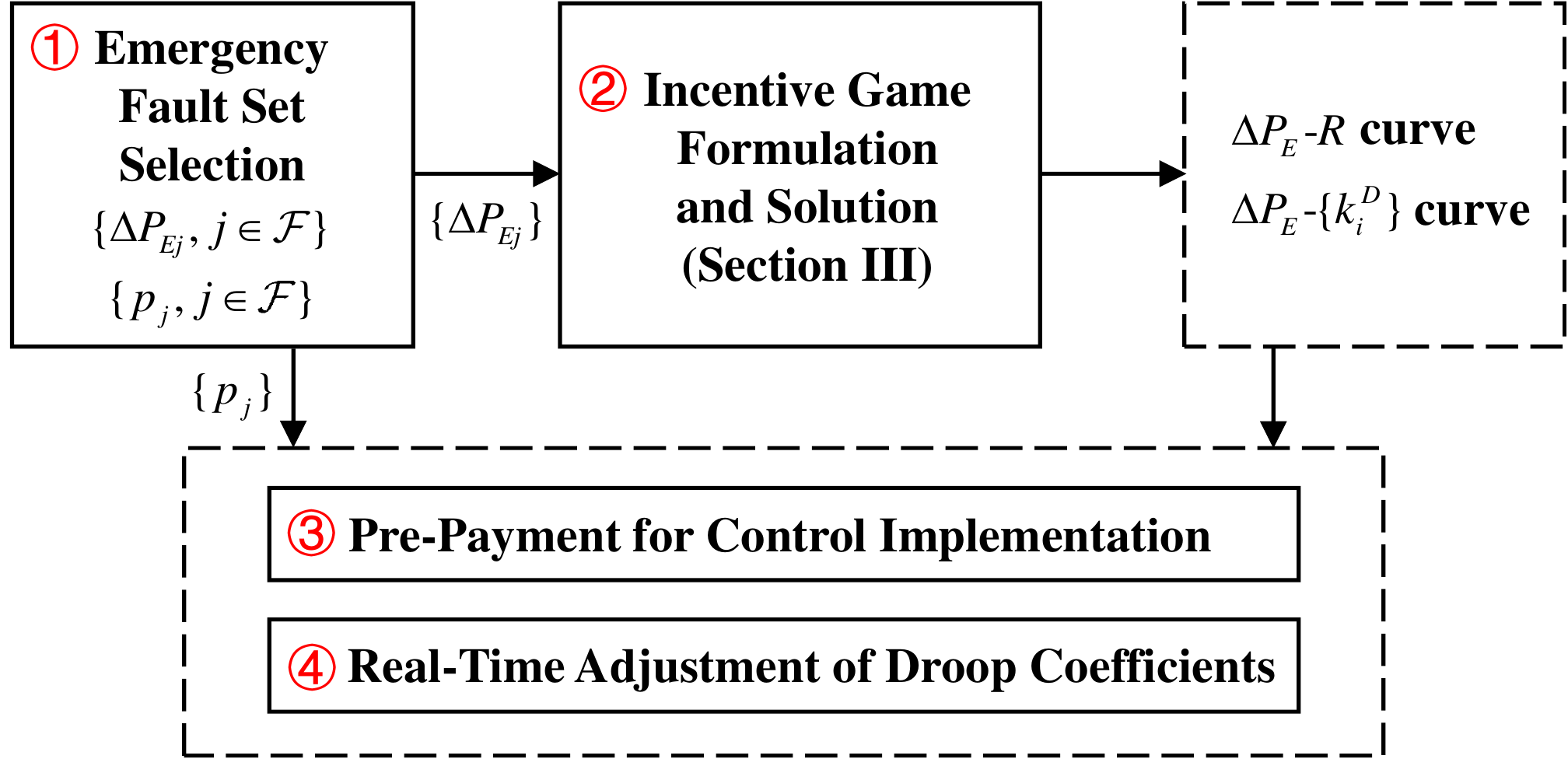}
	\caption{Incentive mechanism for droop-based EFC.}
	\label{incen_mech_flow}
\end{figure}

In this paper, we focus on how to incentivize the AD systems to participate in the EFC and implement the designed LCC-HVDC droop control. Different from the reserve-capacity-based incentive mechanism in the existing literature, a control-parameter-based incentive mechanism is proposed in this section. The main idea of the incentive mechanism is that the AD system with more contribution to the system frequency stability obtains more amount of reward. Since the contribution can be characterized by the droop coefficient, in the proposed incentive mechanism, the AM system provides a total reward $R$ to the AD systems, and the allocated reward to each AD system is proportional to the droop coefficient of its connected LCC-HVDC system. The proposed incentive mechanism for droop-based EFC is shown in Fig. \ref{incen_mech_flow}, which follows these steps below.

\textit{Step 1: Emergency Fault Set Selection}. The AM system first selects the set of emergency faults $\mathcal{F}$ by enumeration approach. The set $\mathcal{F}$ is supposed to contain all the concerned emergency frequency faults, e.g., the HVDC block faults and the generator tripping faults, and the number of faults in $\mathcal{F}$ is denoted by $n_F$. The power imbalances corresponding to the faults in $\mathcal{F}$ are $\{\Delta P_{Ej},j\in\mathcal{F}\}$, where $\Delta P_{Ej}>0$ for power shortage or $\Delta P_{Ej}<0$ for power redundancy. Then, the fault ratios $\{p_{j},j\in\mathcal{F}\}$ (defined as the ratio of the occurrence times of fault $j$ to the total occurrence times of all faults in $\mathcal{F}$) can be estimated via the historical data. We have $\sum_{j\in\mathcal{F}} p_j=1$. The execution cycle $T$ of the designed mechanism is the average time period of one emergency fault occurrence.

\begin{figure}[tb]
	\centering
	%\vspace{-0.2cm}  %调整图片与上文的垂直距离
	%\setlength{\abovecaptionskip}{-0.05cm}   %调整图片标题与图距离
	%\setlength{\belowcaptionskip}{-2cm}   %调整图片标题与下文距离
	\includegraphics[width=0.46\textwidth]{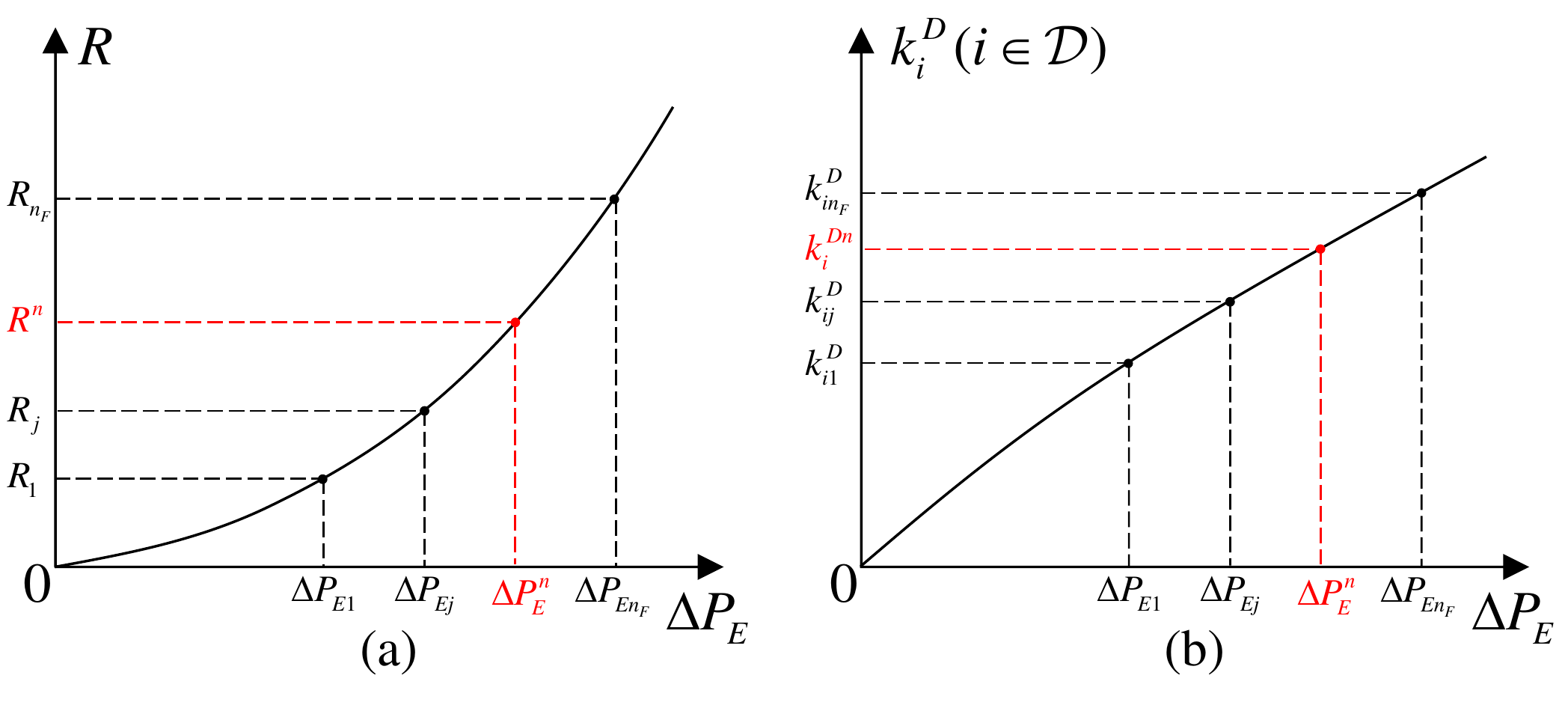}
	\caption{Schematic diagram. (a) $\Delta P_{E}$-$R$ curve. (b) $\Delta P_{E}$-$\{k_i^D\}$ curve.}
	\label{dp_curve}
\end{figure}

\textit{Step 2: Incentive Game Formulation and Solution}. For each $\Delta P_{Ej}$ in $\mathcal{F}$, the corresponding reward $R_j$ for AD systems and the corresponding droop coefficients $\{k_{ij}^D,i\in\mathcal{D}\}$ for LCC-HVDC systems are determined by the Nash equilibrium of a game among AM system and AD systems, which is called the incentive game. The incentive game is organized by a supporting platform. The modeling and solution of the incentive game are introduced in detail in Section III, where the frequency regulation constraints and frequency security constraints are considered in the game modeling, and an algorithm for seeking the equilibrium is proposed. According to the incentive game with each $\Delta P_{Ej}$ in $\mathcal{F}$, the $\Delta P_{E}$-$R$ curve and the $\Delta P_{E}$-$\{k_i^D\}$ curves can be derived, which are composed by a series of scatter points, as shown in Fig. \ref{dp_curve}. Based on the above curves, the following pre-payment and real-time adjustment are carried out.

\textit{Step 3: Pre-Payment for Control Implementation}. Base on the fault ratios $\{p_{j}\}$, we define the expected power imbalance as $\Delta \tilde{P}_E = \sum_{j\in\mathcal{F}} (p_j \Delta P_{Ej})$, and the nearest-to-expected value of power imbalance $\Delta P_E^n$ in $\mathcal{F}$ is defined as the solution of the following optimization problem \eqref{nearest-to-expected}
\begin{align}
\label{nearest-to-expected}
\min_{\Delta P_{Ej}\in \mathcal{F}}\vert \Delta P_{Ej}-\Delta \tilde{P}_E \vert
\end{align}
At the start of the incentive mechanism execution, the AM system pre-pays the reward $R^n$ corresponding to $\Delta P_E^n$ to the AD systems. Then, the AD systems implement the designed droop-base EFC strategy in the LCC-HVDC systems in advance and set their droop coefficients with the $\{k_i^{Dn}\}$ corresponding to $\Delta P_E^n$, and allocate the reward $R^n$ in proportion to the droop coefficients. The $R^n$ and $\{k_i^{Dn}\}$ are shown in Fig. \ref{dp_curve}.

\textit{Step 4: Real-Time Adjustment of Droop Coefficients}. Supposing that an emergency frequency fault with power imbalance $\Delta P_E^r$ occurs during the incentive mechanism execution cycle. As shown in Fig. \ref{dp_curve}, if $\Delta P_E^r \le \Delta P_E^n$, the EFC strategy with the pre-set droop coefficients can satisfy the frequency security constraints and there is no need to adjust the droop coefficients. If $\Delta P_E^r > \Delta P_E^n$, the pre-set droop coefficients cannot meet the frequency security constraints and need real-time adjustments according to the practical $\Delta P_E^r$. There are two ways for the AM system to obtain the $\Delta P_E^r$ in real time: 1) direct fault diagnosis approach to perceive the $\Delta P_E^r$, 2) indirect frequency-measurement-based approach, and in this paper, we adopt the former. Then, the signal of $\Delta P_E^r$ is transmitted to each LCC-HVDC controller, and the droop coefficients are adjusted by looking up the $\Delta P_{E}$-$\{k_i^D\}$ curves.

 \textbf{Remark 1:} Several remarks on the designed incentive mechanism for EFC: 
 
 1) Considering the advantages of the designed mechanism, since the droop coefficients of LCC-HVDC systems can be adjusted immediately according to the practical power imbalance, the proposed incentive mechanism can deal with various possible emergency faults. Moreover, in the proposed mechanism, the short-time overload capability
 of LCC-HVDC systems is utilized to provide power support for EFC. In normal conditions, LCC-HVDC systems have no need to reserve capacity, which ensures the operation economy.
 
 2) Besides the droop-based EFC, the control-parameter-based design idea for incentive mechanism can be extended to other EFC strategies, so long as an index characterizing the controller's contribution to system frequency stability can be defined according to control parameters.
 
 3) Although the reward allocation rule of the proposed mechanism is based on the droop coefficient, we also consider the power regulation margin constraint of the LCC-HVDC system in the following incentive game modeling. Therefore, an LCC-HVDC system with small power regulation margin will not provide a large droop coefficient.  

\section{Incentive Game Formulation and Analysis}

In this section, the incentive game model is formulated to implement the proposed incentive mechanism. Then, an algorithm for seeking the Nash equilibrium is proposed, which can be easily applied to engineering practice. Moreover, the uniqueness of the incentive equilibrium is analyzed.

\subsection{Incentive Game Model}

For each power imbalance $\Delta P_{Ej}$ in $\mathcal{F}$, the incentive game model is formulated to derive the corresponding $R_j$ and $\{k_{ij}^D\}$. In the following model, we omit the subscript $j$ for brevity. In the incentive game, the AM system and the AD systems solve their respective optimization problems to minimize the disutilities. 

For the AM system, the optimization objective consists of two parts: 1) minimizing the reward $R$ to the AD systems, 2) minimizing the total power regulation costs of the generators in the AM system. We define the cost function of generator $i \in \mathcal{G}^{am}$ in a classic form $C_i^{am}(\Delta P_i^G) = \frac{1}{2} \alpha_i (\Delta P_i^G)^2$ \cite{wood2013power}, where $\Delta P_i^G$ is the power regulation, and $\alpha_i$ is the cost coefficient for generator $i$. The optimization problem of the AM system (OAM problem) is shown in \eqref{oam_pro}, where the decision variable is $R$.
\begin{subequations}
	\label{oam_pro}
	\begin{align}
	\min_{R}\ &F^{am}(R) = R+\sum_{i\in \mathcal{G}^{am}} \frac{1}{2} \alpha_i (\Delta P_i^G)^2
	\label{oam_pro_1}\\
	\text{s.t.}\ &\Delta P_{E} = -\Big(\sum_{i\in \mathcal{D}} k_i^D+\sum_{i\in \mathcal{G}^{am}} k_i^G \Big) \hat{\omega}^{am}
	\label{oam_pro_2}\\
	&\Delta P_i^G = -k_i^G \omega^{am},\ i\in \mathcal{G}^{am}
	\label{oam_pro_3}\\
	&R = Y(\omega^{am}-\hat{\omega}^{am})
	\label{oam_pro_4}\\
	&\underline{P}_i^G \le P_i^G+\Delta P_i^G \le	\overline{P}_i^G,\ i\in \mathcal{G}^{am}
	\label{oam_pro_5}\\
	&\underline{\omega}^{am} \le \omega^{am} \le \overline{\omega}^{am}
	\label{oam_pro_6}\\
	&\underline{R} \le R \le \overline{R}
	\label{oam_pro_7}
	\end{align}
\end{subequations}
where $F^{am}(R)$ is the disutility function of the AM system, $k_i^G$ is the droop coefficient of generator $i$, $\hat{\omega}^{am}$ is the calculated value of the frequency deviation according to the given $\{k_i^D\}$, which shows whether the given $\{k_i^D\}$ by the AD systems can satisfy the frequency security constraints, $\omega^{am}$ is the expected steady-state frequency deviation, $Y(\cdot)$ is a response function, $P_i^G$ is the nominal power of generator $i$, $\underline{P}_i^G$, $\underline{\omega}^{am}$, $\underline{R}$ and $\overline{P}_i^G$, $\overline{\omega}^{am}$, $\overline{R}$ are the lower and upper bounds of $P_i^G$, $\omega^{am}$, $R$, respectively. \eqref{oam_pro_2} is the frequency regulation equation of the AM system, \eqref{oam_pro_3} is the expression of $\Delta P_i^G$, \eqref{oam_pro_4} describes the response of the AM system to $(\omega^{am}-\hat{\omega}^{am})$, and in the following subsection, we make an analytical assumption on the response function $Y(\cdot)$, \eqref{oam_pro_6} is the frequency security constraint.

For AD $i$ system, $i \in \mathcal{D}$, the optimization objective also consists of the following two parts: 1) maximizing the shared reward in proportion to $k_i^D$, 2) minimizing the total costs of the generators in AD $i$ system. The shared reward of AD $i$ can be represented as $(\frac{k_i^D}{\sum_{l\in\mathcal{D}} k_l^D}) R$, and the generators in AD $i$ also use the classic cost function as aforementioned. The optimization problem of AD $i$ system (OAD $i$ problem) is as follows, where the decision variable is $k_i^D$.
\begin{subequations}
	\label{oad_pro}
	\begin{align}
	\min_{k_i^D}\ &F^{ad}_i (k_i^D) = -(\frac{k_i^D}{\sum_{l\in\mathcal{D}} k_l^D})R+\sum_{h\in \mathcal{G}^{ad}_i} \frac{1}{2} \alpha_h (\Delta P_h^G)^2
	\label{oad_pro_1}\\
	\text{s.t.}\ &\Delta P_{i}^D = -k_i^D \omega^{am} = -\Big(\sum_{h\in \mathcal{G}_i^{ad}} k_h^G \Big) {\omega}^{ad}_i
	\label{oad_pro_2}\\
	&\Delta P_h^G = -k_h^G \omega^{ad}_i,\ h\in \mathcal{G}^{ad}_i
	\label{oad_pro_3}\\
	&\underline{P}_i^D \le P_i^D+\Delta P_i^D \le	\overline{P}_i^D
	\label{oad_pro_4}\\
	&\underline{P}_h^G \le P_h^G+\Delta P_h^G \le	\overline{P}_h^G,\ h\in \mathcal{G}^{ad}_i
	\label{oad_pro_5}\\
	&\underline{\omega}^{ad}_i \le \omega^{ad}_i \le \overline{\omega}^{ad}_i
	\label{oad_pro_6}
	\end{align}
\end{subequations}
where $F^{ad}_i (k_i^D)$ is the disutility function of AD $i$ system, $\alpha_h$, $k_h^G$, $\Delta P_h^G$ and $P_h^G$ are respectively the cost coefficient, primary droop coefficient, power regulation and nominal power of generator $h \in \mathcal{G}^{ad}_i$, $\omega^{ad}_i$ is the frequency deviation of the AD $i$ system, $\Delta P_i^D$ is the power regulation of LCC-HVDC $i$, $\underline{P}_h^G$, $\underline{\omega}^{ad}_i$ and $\overline{P}_h^G$, $\overline{\omega}^{ad}_i$ are the lower and upper bounds of $P_h^G$ and $\omega^{ad}_i$ respectively, $\underline{P}_i^D$ and $\overline{P}_i^D$ are the lower and upper bounds of $P_i^D$, which are also $>0$ for SE-LCC or $<0$ for RE-LCC. \eqref{oad_pro_2} includes the P-f droop equation of LCC-HVDC $i$ and the frequency regulation equation of AD $i$ system, \eqref{oad_pro_3} expresses $\Delta P_h^D$, and \eqref{oad_pro_6} is the frequency security constraint. 

In the designed mechanism, the process of AM system and AD systems solving the above optimization problems presents a non-cooperative game, i.e., the incentive game. The three key elements of the incentive game \cite{mei2017engineering,mei2012game} are summarized as follows:
\begin{itemize}
	\item \textit{Players}: The AM system and the AD systems.   
	\item \textit{Strategies}: The reward $R \in \Omega^R$ for the AM system, and the droop coefficient $k_i^D \in \Omega^{D}_i$ for AD $i$ system. The $\Omega^R$ and $\Omega^{D}_i$ are the feasible sets of the strategies, which can be determined by the constraints \eqref{oam_pro_5}-\eqref{oam_pro_7} and \eqref{oad_pro_4}-\eqref{oad_pro_6}, respectively.
	\item \textit{Payoffs}: The disutilities $F^{am}(R)$ for the AM system, and $F^{ad}_i (k_i^D)$ for AD $i$ system.
\end{itemize}

We denote the vector $k^D = \{k_i^D,i\in \mathcal{D}\}$, then, the following definition of the Nash equilibrium of the incentive game is given:

\textbf{Definition 1 (Nash Equilibrium):} A strategy profile $x^*=(R^*,k^{D*})$ is a Nash equilibrium of the incentive game if:
\begin{align}
&F^{ad}_i (R^*,k^{D*}) \le F^{ad}_i (R^*,k_i^D,k_{-i}^{D*}),\ \forall k_i^D \in \Omega^{D}_i,\ i\in \mathcal{D}
\nonumber\\
&F^{am}(R^*,k^{D*}) \le F^{am}(R,k^{D*}),\ \forall R \in \Omega^R
\label{nash_equ}
\end{align}

Subsequently, an algorithm for seeking the equilibrium of the incentive game is designed, which can be easily applied to engineering practice. 
 
\subsection {Algorithm for Seeking the Equilibrium}

In this subsection, a simple but effective approach is utilized to design the algorithm for seeking the incentive equilibrium, i.e., the fixed point method \cite{agarwal2001fixed}, in which the players of the game iteratively solve their respective optimization problems to find the equilibrium. When one player solves its optimization problem, the decision variables of the other players are regarded as given constants. Nevertheless, due to the couplings among the decision variables $\{k_i^D\}$ in the first items of \eqref{oad_pro_1} in the OAD problems, seeking the Nash equilibrium directly by the fixed point method might result in problems such as initial value dependence and decision order dependence \cite{wang2021distributed}. Therefore, a virtual price variable $\gamma$ is introduced to eliminate the aforementioned couplings, which is define as:
\begin{align}
\label{virtual_price}
\gamma = \frac{R}{\sum_{l\in\mathcal{D}} k_l^D}
\end{align}
When $k^D$ is given by the AD systems, the reward $R$ can be determined independently by $\gamma$, thus, in the algorithm design, we regard $\gamma$ as the decision variable of the AM system in place of $R$. 

Then, based on the defined $r$, a linear response function $\hat{Y}(\cdot)$ is adopted to describe the constraint \eqref{oam_pro_4} in the OAM problem, which is shown as follows:
\begin{align}
\label{respones_func}
\gamma = \hat{Y}(\omega^{am}-\hat{\omega}^{am})=a(\omega^{am}-\hat{\omega}^{am})+\gamma^{t-1}
\end{align}
where $a$ is the marginal response coefficient with $\underline{a} \le a \le \overline{a}$, $\gamma^{t-1}$ is the virtual price value from the last fixed-point iteration. In engineering practice, since the AM system can determine the expected steady-state frequency after emergency faults according to the technical guide for power system security and stability, the expected steady-state frequency deviation $\omega^{am}$ is assumed to be a constant given by the AM system, and the impact of $\omega^{am}$ on the Nash equilibrium is analyzed in the case study. \eqref{respones_func} describes the decision-making willingness of the AM system, i.e., when the calculated frequency deviation $\hat{\omega}^{am}$ is different from $\omega^{am}$, the AM system will adjust the virtual price $\gamma$ according to \eqref{respones_func}.

Substitute \eqref{virtual_price}-\eqref{respones_func} into \eqref{oam_pro}, and remove constraints \eqref{oam_pro_3}, \eqref{oam_pro_5}-\eqref{oam_pro_6} and the second item of \eqref{oam_pro_1} since $\omega^{am}$ is assumed to be a constant, then, the modified OAM (MOAM) problem in the compact form is shown in \eqref{moam_pro}:
\begin{subequations}
	\label{moam_pro}
	\begin{align}
	\min_{\gamma}\ &\hat{F}^{am}(\gamma) = \Big(\sum_{i\in\mathcal{D}} k_i^D \Big) \gamma
	\label{moam_pro_1}\\
	\text{s.t.}\ &\Delta P_{E} = -\Big(\sum_{i\in \mathcal{D}} k_i^D+\sum_{i\in \mathcal{G}^{am}} k_i^G \Big) \hat{\omega}^{am}
	\label{moam_pro_2}\\
	&\gamma = a(\omega^{am}-\hat{\omega}^{am})+\gamma^{t-1}
	\label{moam_pro_3}\\
	&\gamma \in \Omega^{\gamma}
	\label{moam_pro_4}
	\end{align}
\end{subequations}
where $\Omega^{\gamma}$ is the feasible set of $\gamma$ which can be derived by \eqref{oam_pro_7} and $\underline{a} \le a \le \overline{a}$. Substitute \eqref{virtual_price} and \eqref{oad_pro_2}-\eqref{oad_pro_3} into \eqref{oad_pro_1}, and the modified OAD (MOAD) $i$ problem in the compact form is shown in \eqref{moad_pro}:
\begin{subequations}
	\label{moad_pro}
	\begin{align}
	\min_{k_i^D}\ &\hat{F}^{ad}_i (k_i^D) \nonumber\\&= -\gamma {k_i^D}+\frac{(\omega^{am})^2 \sum_{h\in \mathcal{G}^{ad}_i} \frac{1}{2} \alpha_h (k_h^G)^2}{\Big( \sum_{h\in \mathcal{G}^{ad}_i} k_h^G \Big)^2} (k_i^D)^2
	\label{moad_pro_1}\\
	\text{s.t.}\ &k_i^D \in \Omega_i^D
	\label{moad_pro_2}
	\end{align}
\end{subequations}
where the feasible set $\Omega_i^D = \{k_i^D|\underline{k}_i^D \le k_i^D \le \overline{k}_i^D\}$ is defined in Section III.A. In the algorithm for seeking the equilibrium with the fix point method (SE-FP algorithm), the modified Nash equilibrium $\hat{x}^* = (\gamma^*,k^{D*})$ is first derived by the fixed point approach. Then, the Nash equilibrium $x^*=(R^*,k^{D*})$ can be obtained by \eqref{virtual_price}. The details of the algorithm for seeking the equilibrium is shown in Algorithm 1. 
\begin{algorithm}[tb]
	\caption{SE-FP Algorithm} 
	\label{alg_seek}
	
	\hangafter 1
	\hangindent 1em
	\textbf{Input:} Power imbalance $\Delta P_E$, initial value of virtual price $\gamma^0$, initial value of droop coefficients $k^{D,0} = \{k_i^{D,0},i\in \mathcal{D}\}$, expected frequency deviation $\omega^{am}$, very small error tolerances $\varepsilon_\gamma, \varepsilon_k>0$, and iteration index $t=1$.
	
	\hangafter 1
	\hangindent 1em
	\textbf{Output:} Nash equilibrium of the incentive game $x^*=(R^*,k^{D*})$.
	
	\hangafter 1
	\hangindent 1em
	\textbf{Step 1:} Given $\gamma^{t-1}$ and $\omega^{am}$, for each $i\in \mathcal{D}$, derive $k_i^{D,t}$ by solving the MOAD $i$ problem \eqref{moad_pro}, respectively.
	
	\hangafter 1
	\hangindent 1em
	\textbf{Step 2:} Given $k^{D,t}=\{k_i^{D,t}\}$ and $\omega^{am}$, derive $\gamma^t$ by solving the MOAM problem \eqref{moam_pro}.
	
	\hangafter 1
	\hangindent 1em
	\textbf{Step 3:} Calculate the errors $e_\gamma^t = \gamma^t - \gamma^{t-1}$ and the error vector $e_k^t = k^{D,t} - k^{D,t-1}=\{e_{ki}^t, i\in \mathcal{D}\}$.
	
	\hangafter 1
	\hangindent 1em
	\textbf{If} $| e_\gamma^t | < \varepsilon_\gamma$ and $\max\{|e_{ki}^t|,i\in \mathcal{D}\} < \varepsilon_k$, the fixed point method converges, continue \textbf{Step 4}.
	
	\hangafter 1
	\hangindent 1em
	\textbf{else} Set $t=t+1$. Return to \textbf{Step 1}. 
	
	\hangafter 1
	\hangindent 1em
	\textbf{Step 4:} Derive the modified Nash equilibrium and the Nash equilibrium by \eqref{algo_nash}. \textbf{exit}.
	\begin{subequations}
	\label{algo_nash}
	\begin{align}
	&\hat{x}^* = (\gamma^*,k^{D*}) = (\gamma^t,k^{D,t})\\
	&x^*=(R^*,k^{D*}) = ((\sum_{i\in\mathcal{D}} k_i^{D*})\gamma^*,k^{D*})
	\end{align}
	\end{subequations}  
\end{algorithm}

\begin{figure}[tb]
	\centering
	%\vspace{-0.2cm}  %调整图片与上文的垂直距离
	%\setlength{\abovecaptionskip}{-0.05cm}   %调整图片标题与图距离
	%\setlength{\belowcaptionskip}{-2cm}   %调整图片标题与下文距离
	\includegraphics[width=0.39\textwidth]{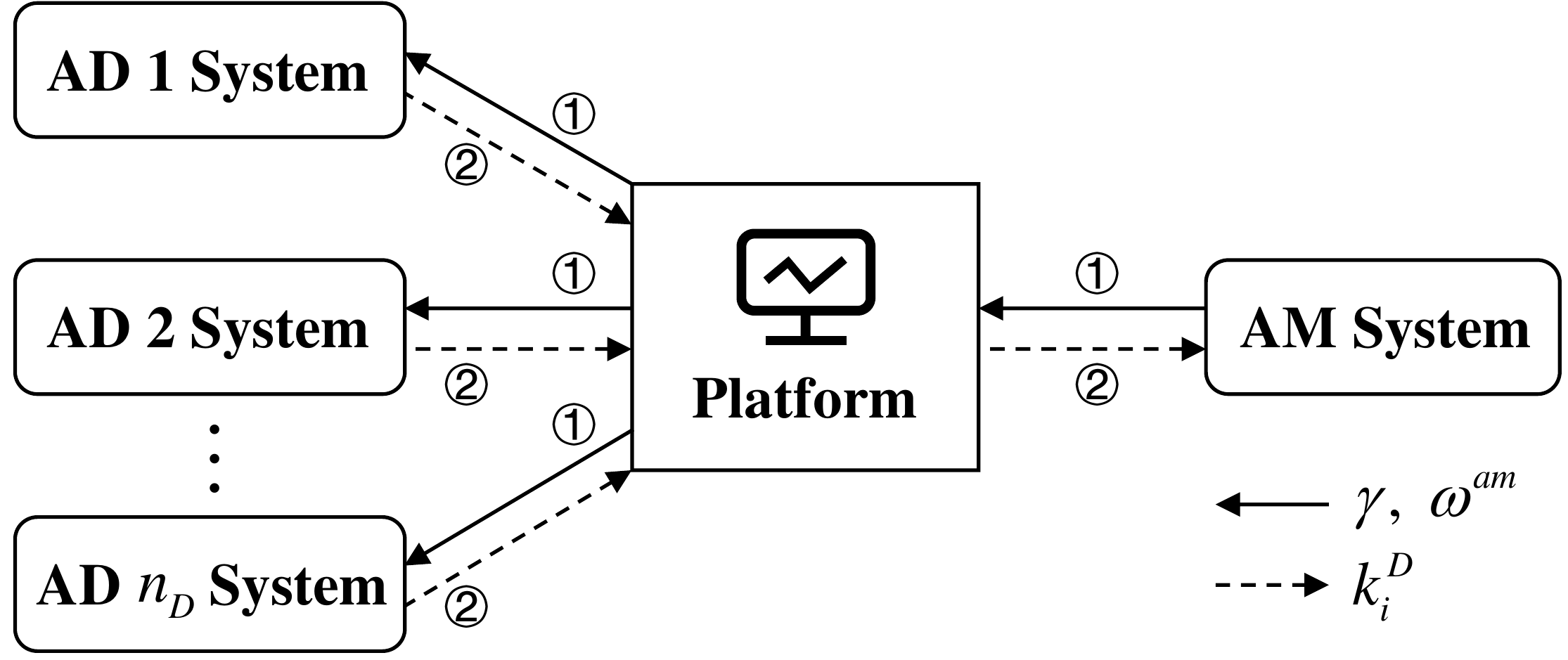}
	\caption{Incentive game framework among AM and AD systems.}
	\label{frame_game}
\end{figure}

Considering the implementation of the formulated incentive game with the SE-FP algorithm, Fig. \ref{frame_game} shows the framework of the incentive game, where the AM system and the AD systems realize information transmissions through a platform. In Fig. \ref{frame_game}, only a small amount of information (i.e., the decision variables $\gamma$ and $k^D_i$, and given constant $\omega^{am}$) needs to be transmitted, thus, the incentive game with the SE-FP algorithm can be easily implemented in practical projects, and the private information (e.g., the cost coefficients and the droop coefficients of the generators) of the players can be preserved. Moreover, the convergence of the SE-PF algorithm is analyzed in the case study, which can be guaranteed with different initial values.  

\subsection{Uniqueness of the Equilibrium}

In this subsection, the uniqueness of the Nash equilibrium of the incentive game is analyzed, which ensures that there will not appear multiple equilibriums when applying the proposed incentive mechanism. We make the following assumptions.

\textbf{Assumption 3:} For the incentive game, there exists an Nash equilibrium $x^*=(R^*,k^{D*})$, $R^* \in \Omega^R$, $k_i^{D*} \in \Omega^{D}_i$.

\textbf{Assumption 4:} The droop coefficients of the LCC-HVDC systems satisfy $\sum_{i\in \mathcal{D}} \underline{k}_i^D <\sum_{i\in \mathcal{D}} k_i^D <\sum_{i\in \mathcal{D}} \overline{k}_i^D$. 

In practice, Assumption 4 is usually considered as a security constraint which implies that the sum of the droop coefficients will not reach the lower and upper bounds. Then, we have the following proposition for the uniqueness of the equilibrium.

\textbf{Proposition 1:} If Assumption 3-4 hold, the Nash equilibrium $x^*=(R^*,k^{D*})$ of the incentive game is unique.

\begin{proof}
The Nash equilibrium $x^*$ is unique if and only if the modified Nash equilibrium $\hat{x}^* = (\gamma^*,k^{D*})$ is unique, thus, we just need to analyze the uniqueness of $\hat{x}^*$. 

Before the uniqueness analysis of the equilibrium, we first analyze the monotonicity of the optimal decision $k_i^{D,opt}$ with respect to the virtual price $\gamma$. According to the optimality condition (KKT condition) \cite{boyd2004convex}, the optimal decision of the MOAD $i$ problem is:
\begin{align}
\label{opt_moad}
k_i^{D,opt} = \left[ \frac{\Big( \sum_{h\in \mathcal{G}^{ad}_i} k_h^G \Big)^2 \gamma}{2 (\omega^{am})^2 \sum_{h\in \mathcal{G}^{ad}_i} \frac{1}{2} \alpha_h (k_h^G)^2} \right]_{\Omega^D_i}
\end{align}
where the $[\cdot]_{\Omega^D_i}$ represents the projection on $\Omega^D_i$. According to \eqref{opt_moad}, for each $i\in\mathcal{D}$, $k_i^{D,opt}$ is monotonically increasing with respect to $\gamma$ when $k_i^{D,opt}$ is an interior point of $\Omega_i^D$, and $k_i^{D,opt}$ tends to a saturated value $\overline{k}^D_i$ when $\gamma$ continuously increases. Then, if Assumption 4 holds, there exists at least one AD $i$ system with $\underline{k}_i^D < k_i^{D,opt} < \overline{k}_i^D$, thus, the sum $\sum_{i\in \mathcal{D}} k_i^{D,opt}$ is monotonically increasing with respect to the virtual price $\gamma$.

Based on the monotonicity, we further analyze the uniqueness of the modified Nash equilibrium. According to the modified incentive game with the SE-FP algorithm, one of the conditions at the equilibrium is:
\begin{align}
\label{con_equ}
|\gamma^* - \gamma^{t-1}|<\varepsilon_\gamma
\end{align} 
where $\varepsilon \approx 0$ is a very small number. Then, combining \eqref{moam_pro_3} and \eqref{con_equ}, we have $\hat{\omega}^{am*} = \omega^{am}$, which yields the following so-called equilibrium constraint:
\begin{align}
\label{equ_cons}
\sum_{i\in \mathcal{D}} k_i^{D*} = -\frac{\Delta P_{E}}{\omega^{am}}- \sum_{i\in \mathcal{G}^{am}} k_i^G
\end{align}
If Assumption 3 holds, there exists a $k^{D*}$ satisfying \eqref{equ_cons}. From the monotonicity, there exists at most one virtual price $\gamma^*$ which is corresponding to the $k^{D*}$. Therefore, the modified Nash equilibrium $\hat{x}^* = (\gamma^*,k^{D*})$ is unique, which completes the proof.
\end{proof}

\textbf{Remark 2:} In engineering practice, if the power imbalance is too large, all the droop coefficients might reach their saturated values (upper bounds), and the Assumption 3-4 cannot hold in the incentive game. In fact, at this time, the LCC-HVDC systems and generators cannot provide enough power support to cover the emergency power imbalance. To deal with this scenario, the incentive game with the SE-FP algorithm can give the minimal reward $R$ corresponding to the saturated droop coefficients, and the uncovered part of the power imbalance needs to be undertaken coordinately by the load shedding operations.

\section{Properties of the Incentive Mechanism}
 
In this section, several preferable properties of the designed incentive mechanism for EFC are analyzed. In this mechanism, the pre-payments and the pre-settings of the droop coefficients are all based on the Nash equilibrium of the incentive game with the nearest-to-expected power imbalance, thus, we show the properties of the incentive mechanism by analyzing the properties of the incentive game model. 

\subsection{Individual Rationality and Incentive Compatibility}

The individual rationality means that AD systems can obtain more utilities by participating in the designed incentive mechanism than not, thus, they are willing to participate in the mechanism \cite{nisan2001algorithmic}. Considering the disutility function $\hat{F}^{ad}_i (k_i^D)$ of the AD $i$ system in the incentive game model, when AD $i$ does not participate in the incentive game (i.e., $k_i^D=0$), we have $\hat{F}^{ad}_i (0) = 0$ according to \eqref{moad_pro_1}. Then, we have the following proposition about the individual rationality of the incentive game.

\textbf{Proposition 2:} The incentive game is individual rational for AD systems, i.e., given $\gamma>0$, the disutility with the optimal decision satisfies $\hat{F}^{ad}_i (k_i^{D,opt})<0$.
\begin{proof}
Obtain the derivative of $\hat{F}^{ad}_i (k_i^D)$ with respect to $k_i^D$, which is:
\begin{align}
\frac{d\hat{F}^{ad}_i (k_i^D)}{d k_i^D}= -\gamma+\frac{2(\omega^{am})^2 \sum_{h\in \mathcal{G}^{ad}_i} \frac{1}{2} \alpha_h (k_h^G)^2}{\Big( \sum_{h\in \mathcal{G}^{ad}_i} k_h^G \Big)^2} k_i^D
\end{align}
Since $0 \in \Omega_i^D$, according to \eqref{opt_moad}, we have:
\begin{align}
\frac{d\hat{F}^{ad}_i (k_i^D)}{d k_i^D}<0,\ \forall k_i^D\in (0,k_i^{D,opt})
\end{align}
Thus, $\hat{F}^{ad}_i (k_i^D)$ is monotonically decreasing in the interval $(0,k_i^{D,opt})$, which yields:
\begin{align}
\hat{F}^{ad}_i (k_i^{D,opt})<\hat{F}^{ad}_i (0) =0
\end{align}
which completes the proof.	
\end{proof}

The individual rationality analysis shows that the designed incentive mechanism is reasonable and can provide positive incentives for the AD systems to participate in the game.

Moreover, the incentive compatibility \cite{fang2022efficient} in the proposed incentive mechanism for EFC means that each AD system is willing to give the true optimal droop coefficient instead of a false one, i.e., giving the true droop coefficient will obtain more utility than not. In fact, the incentive compatibility can be satisfied since the incentive mechanism is based on the Nash equilibrium of the incentive game. According to Definition 1, when at the Nash equilibrium, there is no player being able to obtain more utility by changing its own strategy, i.e., 
\begin{align}
F^{ad}_i (R^*,k^{D*}) \le F^{ad}_i (R^*,k_i^{D,f},k_{-i}^{D*}),\ \forall k_i^{D,f} \in \Omega^{D}_i
\end{align}  
where $k_i^{D,f}$ is a false droop coefficient. Therefore, the AD systems are willing to give the true droop coefficients by optimal decision-making, which guarantees the incentive compatibility.
%In the designed incentive mechanism for EFC, since the given reward of one AD system is proportional to its set droop coefficient, one AD system might give a false larger droop coefficient to obtain a larger share of the reward. For the incentive game in this paper,  

\subsection{Social Optimality}

To analyze the social efficiency of the proposed incentive mechanism, we first define the social welfare problem \cite{chen2019energy}. When an emergency fault occurs and the droop-based EFC strategy works, the optimization objective of the social planner is minimizing the total power regulation costs of the generators in all the AD systems. Since the social welfare problem is considered from the perspective of the whole power system, we ignore the reward $R$ in this problem. The social welfare problem can be represented as \eqref{social_opt}.
\begin{subequations}
	\label{social_opt}
	\begin{align}
	\min_{\{k_i^D\}}\ &\sum_{i\in \mathcal{D}}\sum_{h\in \mathcal{G}^{ad}_i} \frac{1}{2} \alpha_h (\Delta P_h^G)^2
	\label{social_opt_1}\\
	\text{s.t.}\ &\Delta P_{E} = -\Big(\sum_{i\in \mathcal{D}} k_i^D+\sum_{i\in \mathcal{G}^{am}} k_i^G \Big) \omega^{am}
	\label{social_opt_2}\\
	&\Delta P_{i}^D = -k_i^D \omega^{am} = -\Big(\sum_{h\in \mathcal{G}_i^{ad}} k_h^G \Big) {\omega}^{ad}_i,\ i\in \mathcal{D}
	\label{social_opt_3}\\
	&\Delta P_h^G = -k_h^G \omega^{ad}_i,\ h\in \mathcal{G}^{ad}_i,\ i\in \mathcal{D}
	\label{social_opt_4}\\
	&k_i^D \in \Omega_i^D,\ i\in \mathcal{D}
	\label{social_opt_5}
	\end{align}
\end{subequations}

\textbf{Definition 2 (Social Optimum):} The droop coefficients $\widetilde{k}^D=\{\widetilde{k}_i^D, i\in \mathcal{D}\}$ is the social optimum if $\widetilde{k}^D$ is the optimal solution of \eqref{social_opt}.

Then, we have the following proposition about the social efficiency of the incentive game. 

\textbf{Proposition 3:} If Assumption 3-4 holds, for the incentive game and the social welfare problem, the following statements hold:
\begin{enumerate}
	\item the $k^{D*}$ at the Nash equilibrium of the incentive game is identical to the social optimum $\widetilde{k}^D$.
	\item the virtual price $\gamma^*$ at the Nash equilibrium is equal to the opposite value of the Lagrangian multiplier of constraint \eqref{social_opt_2} at the social optimum.
	\item if for $\forall i\in \mathcal{D}$, $\widetilde{k}_i^D$ is an interior point of $\Omega_i^D$, the social optimum $\widetilde{k}^D$ (i.e., the $k^{D*}$) can be derived analytically.
\end{enumerate}
\begin{proof}
	We first simplify the social welfare problem. Substitute \eqref{social_opt_3}-\eqref{social_opt_4} into \eqref{social_opt_1}, the objective of the social welfare problem is represented as:
	\begin{align}
	\sum_{i\in \mathcal{D}} \frac{(\omega^{am})^2 \sum_{h\in \mathcal{G}^{ad}_i} \frac{1}{2} \alpha_h (k_h^G)^2}{\Big( \sum_{h\in \mathcal{G}^{ad}_i} k_h^G \Big)^2} (k_i^D)^2
	\label{social_obj}
	\end{align}
	Let:
	\begin{subequations}
		\label{variable_sub}
		\begin{align}
		&u_i = \frac{(\omega^{am})^2 \sum_{h\in \mathcal{G}^{ad}_i} \frac{1}{2} \alpha_h (k_h^G)^2}{\Big( \sum_{h\in \mathcal{G}^{ad}_i} k_h^G \Big)^2}\\
		&W = -\frac{\Delta P_{E}}{\omega^{am}}- \sum_{i\in \mathcal{G}^{am}} k_i^G
		\end{align}
	\end{subequations}
	We have $u_i>0$, $W>0$. Then, the social welfare problem is simplified as \eqref{social_opt_sim}:
	\begin{subequations}
		\label{social_opt_sim}
		\begin{align}
		\min_{\{k_i^D\}}\ &\sum_{i\in \mathcal{D}} u_i (k_i^D)^2
		\label{social_opt_sim_1}\\
		\text{s.t.}\ &\sum_{i\in \mathcal{D}} k_i^D = W,\ \lambda
		\label{social_opt_sim_2}\\
		&k_i^D \in \Omega_i^D,\ i\in \mathcal{D}
		\label{social_opt_sim_3}
		\end{align}
	\end{subequations}
	where $\lambda$ is the Lagrangian multiplier of constraint \eqref{social_opt_sim_2}.
	
	Considering the conditions satisfied at the modified Nash equilibrium $\hat{x}^* = (\gamma^*,k^{D*})$, $(\gamma^*,k_i^{D*})$ is the optimal solution of the MOAD $i$ problem \eqref{moad_pro}, which satisfies the following KKT condition \cite{ruszczynski2011nonlinear} (with the variable substitutions \eqref{variable_sub}):
	\begin{align}
	0 \in 2 u_i k_i^{D*}-\gamma^*+N_{\Omega_i^D}(k_i^{D*}),\ i\in\mathcal{D}
	\label{kkt_con}
	\end{align}
	Moreover, $(\gamma^*,k^{D*})$ satisfies the following condition according to \eqref{equ_cons}:
	\begin{align}
	\label{equ_con}
	\sum_{i\in \mathcal{D}} k_i^{D*} = W
	\end{align}
	Note that if we identify the Lagrangian multiplier $\lambda$ of constraint \eqref{social_opt_sim_2} with $-\gamma$, the combination of \eqref{kkt_con} and \eqref{equ_con} is also the KKT condition of the social welfare problem \eqref{social_opt_sim}, which means that $k^{D*}$ is identical to $\widetilde{k}^D$ and $\gamma^*$ is equal to $-\widetilde{\lambda}$, where $\widetilde{\lambda}$ is the value corresponding to $\widetilde{k}^D$.
	
	If $\widetilde{k}_i^D = k_i^{D*}$ is an interior point of $\Omega_i^D$, $\forall i\in \mathcal{D}$, we have $\{0\} \in N_{\Omega_i^D}(k_i^{D*})$. Then, by \eqref{kkt_con} and \eqref{equ_con}, the unique $\widetilde{k}^D$ and the modified Nash equilibrium $\hat{x}^*$ can be derived analytically, as shown in \eqref{analy_solu}.
	\begin{subequations}
		\label{analy_solu}
		\begin{align}
		&\widetilde{k}_i^D = k_i^{D*} = \frac{W}{u_i \sum_{i\in\mathcal{D}}(1/u_i)},\ i\in\mathcal{D}\\
		&\widetilde{\lambda} = -\gamma^* = -\frac{2W}{\sum_{i\in\mathcal{D}}(1/u_i)}
		\end{align}
	\end{subequations} 
\end{proof}

\textbf{Remark 3:} Based on the above analysis, the Nash equilibrium under the incentive mechanism satisfies the social optimality, which shows a preferable property. According to Proposition 3, when all the parameters of the entire system are available, seeking the Nash equilibrium of the incentive game is equivalent to solving a convex social welfare problem, which also verifies the uniqueness of the Nash equilibrium. Moreover, in the scenario where the entire power system is operated by a single decision-maker, since the social optimality is satisfied, the incentive mechanism proposed in this paper can also provide reference on how to set the droop coefficients and how to reward the AD systems participating in the EFC strategy.

\section{Case Study}

In this section, the effectiveness of the incentive game is illustrated, and the properties of the proposed mechanism are verified by a case study. Moreover, the impact of the expected frequency deviation $\omega^{am}$ is discussed.   

\subsection{System Description}

The topology of the MIDC test system is shown in Fig. \ref{midc_topo}. In the test system, the AM system contains eight generators, and is connected with four LCC-HVDC systems, in which LCC1, LCC2, LCC3 are SE-LCC systems and LCC4 is an RE-LCC system. There are four AD systems corresponding to the four LCC-HVDC systems, and each AD system contains three equivalent generators. Considering the incentive game model, the related parameters of the generators in the AM system are shown in Table I, and we set the generation cost per unit active power of G1 as the base-value of costs and disutilities. The related parameters of the LCC-HVDC systems and the generators in corresponding AD systems are shown in Table II. The above parameters of the test system are partly from the 4-HVDC modified IEEE New England system in \cite{liu2021optimal,liu2018modeling}. We set the upper and lower bounds of $\omega_i^{ad}$ with $\overline{\omega}_i^{ad}=$0.2 Hz and $\underline{\omega}_i^{ad}=-$0.2 Hz. We set the upper and lower bounds of the marginal response coefficient with $\overline{a}=$20 p.u./MW and $\underline{a}=$10 p.u./MW. The expected frequency deviation is set with $\omega^{am}=-$0.2 Hz for emergency faults with power shortages, and the impact of $\omega^{am}$ is analyzed in Section V.D. 

\begin{table}[tbp] 
	\footnotesize
	%\scriptsize
	\centering
	\vspace{-0.1cm}  %
	\setlength{\abovecaptionskip}{0.cm}
	\setlength{\belowcaptionskip}{-0.cm}
	\caption{Related Parameters of Generators in AM System}
	\label{tab1} 
	\begin{tabular}{ccccc} 
		\toprule 
		No. & $P_i^G$(MW) & $\overline{P}_i^G$, $\underline{P}_i^G$(MW) & $\alpha_i$(p.u./MW$^2$) & $k_i^G$(MW/Hz)\\ 
		\midrule 
		G1 & 320 & 500, 220 & 1.0 & 100\\
		G2 & 350 & 500, 250 & 1.0 & 105\\
		G3 & 370 & 500, 270 & 0.8 & 120\\
		G4 & 380 & 500, 280 & 1.0 & 120\\
		G5 & 400 & 600, 300 & 0.8 & 125\\
		G6 & 425 & 600, 325 & 1.0 & 145\\
		G7 & 450 & 600, 350 & 0.8 & 130\\
		G8 & 470 & 600, 370 & 0.8 & 150\\  
		\bottomrule 
	\end{tabular} 
\end{table}

\begin{table*}[tbp] 
	\footnotesize
	%\scriptsize
	\centering
	\vspace{-0.1cm}  %
	\setlength{\abovecaptionskip}{0.cm}
	\setlength{\belowcaptionskip}{-0.cm}
	\caption{Related Parameters of LCC-HVDC Systems and Generators in AD Systems}
	\label{tab2} 
	\begin{tabular}{cccccccc} 
		\toprule 
		No. & $P_i^D$(MW) & $\overline{P}_i^D$, $\underline{P}_i^D$(MW) & $\{P_h^G\}$(MW) & $\{\overline{P}_h^G\}$(MW) & $\{\underline{P}_h^G\}$(MW) & $\{\alpha_h\}$(p.u./MW$^2$) & $\{k_h^G\}$(MW/Hz)\\ 
		\midrule 
		LCC1/AD1 & 645 & 750, 550 & (610, 540, 650) & (700, 650, 750) & (500, 450, 550) & (0.9, 1.0, 1.1) & (130, 100, 150)\\
		LCC2/AD2 & 630 & 750, 550 & (600, 550, 620) & (700, 650, 750) & (500, 450, 550) & (1.0, 1.0, 0.8) & (155, 120, 140)\\
		LCC3/AD3 & 660 & 750, 550 & (580, 530, 630) & (700, 650, 750) & (500, 450, 550) & (0.9, 1.0, 0.8) & (150, 115, 150)\\
		LCC4/AD4 & 500 & 600, 400 & (590, 560, 640) & (700, 650, 750) & (500, 450, 550) & (1.1, 1.0, 0.8) & (140, 110, 145)\\  
		\bottomrule 
	\end{tabular} 
\end{table*}

In this case study, we select all the single generator tripping faults in the AM system to compose the emergency fault set $\mathcal{F}$, i.e., $\mathcal{F}=\{$F1, $\cdots$, F8$\}$, where the power imbalance $\Delta P_{Ei}$ of F$i$ is corresponding to the nominal power $P_i^G$ of generator $i$ in Table I. When solving the incentive game with $\Delta P_{Ei}$, the corresponding droop coefficient $k_i^G$ of the tripping generator is set to be zero. Then, based on the above settings, we present the following results.

\subsection{Effectiveness of the Incentive Game}

\begin{figure}[tb]
	\centering
	%\vspace{-0.2cm}  %调整图片与上文的垂直距离
	%\setlength{\abovecaptionskip}{-0.05cm}   %调整图片标题与图距离
	%\setlength{\belowcaptionskip}{-2cm}   %调整图片标题与下文距离
	\includegraphics[width=0.49\textwidth]{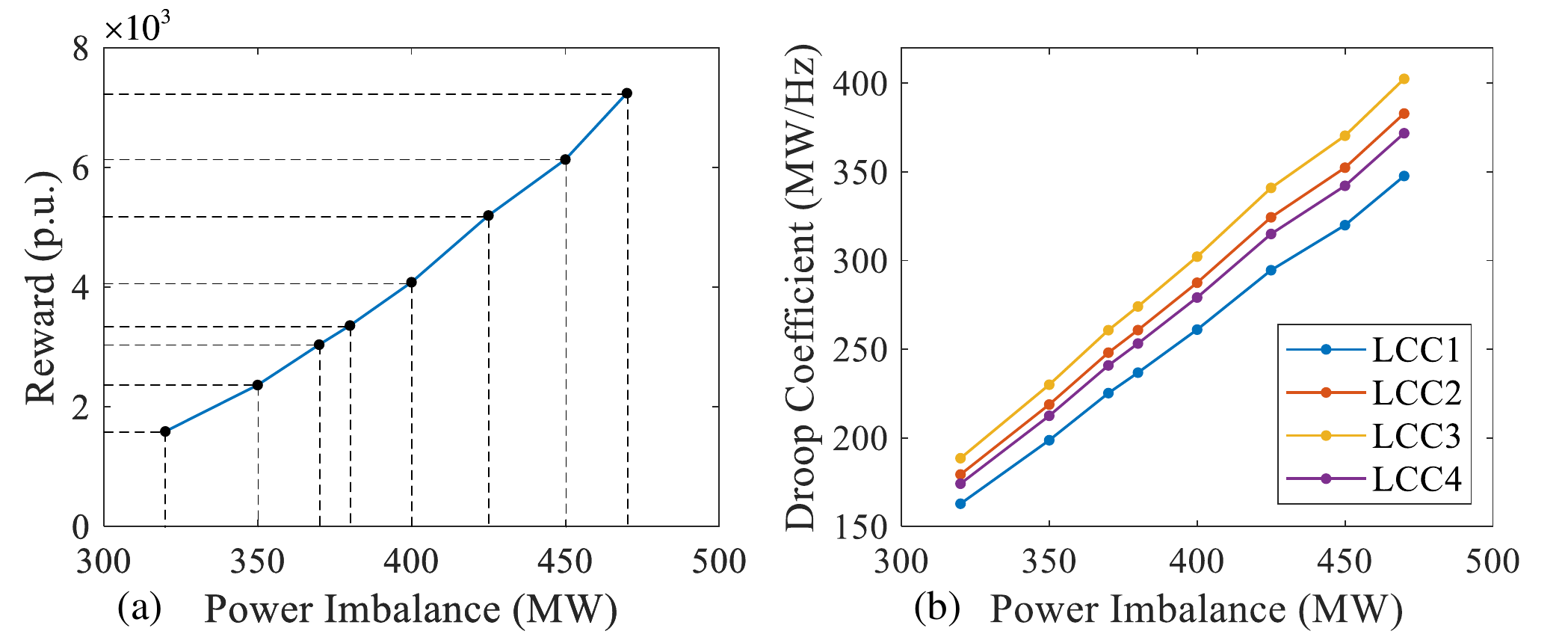}
	\caption{Curves for test system. (a) $\Delta P_{E}$-$R$ curve. (b) $\Delta P_{E}$-$\{k_i^D\}$ curve.}
	\label{case1}
\end{figure}

First, for each emergency fault in $\mathcal{F}$, we derive the Nash equilibrium $x^*=(R^*,k^{D*})$ by the proposed SE-FP algorithm. The Nash equilibriums corresponding to emergency faults are shown in Table III. It can be seen from Table III that for each fault in the selected fault set, the SE-FP algorithm can always seek an equilibrium of the incentive game, which illustrates the effectiveness of the incentive game model and the SE-FP algorithm. Then, according to the Nash equilibriums, the $\Delta P_{E}$-$R$ curve and the $\Delta P_{E}$-$\{k_i^D\}$ curves for the test system can be derived, as shown in Fig. \ref{case1}. Further, according to the proposed incentive mechanism, given the fault ratios $\{p_i,i\in \mathcal{F}\}$, the pre-payment and pre-settings of droop coefficients can be carried out. Supposing that the fault ratios in the fault set of the test system are equal, i.e., $p_i=$0.125, $i\in\mathcal{F}$, from \eqref{nearest-to-expected}, the nearest-to-expected power imbalance is $\Delta P_{E5}=400$ MW, and the pre-payment and pre-settings are carried out according to the Nash equilibrium corresponding to the fault F5. In the incentive mechanism execution cycle, by looking up the curves in Fig. \ref{case1}, the droop coefficients can be adjusted in real time according to the practical emergency fault.

\begin{table*}[tbp] 
	\footnotesize
	%\scriptsize
	\centering
	\vspace{-0.1cm}  %
	\setlength{\abovecaptionskip}{0.cm}
	\setlength{\belowcaptionskip}{-0.cm}
	\caption{Nash Equilibriums Corresponding to Emergency Faults}
	\label{tab3} 
	\begin{tabular}{cccccccc} 
		\toprule 
		No. & $\Delta P_E$(MW) & $\gamma^*$(p.u.$\cdot$Hz/MW) & $k_1^{D*}$(MW/Hz) & $k_2^{D*}$(MW/Hz) & $k_3^{D*}$(MW/Hz) & $k_4^{D*}$(MW/Hz) & $R^*$(p.u.) \\  
		\midrule
		F1 & 320 & 2.25 & 162.88 & 179.38 & 188.55 & 174.18 & 1589.17 \\
		F2 & 350 & 2.75 & 198.69 & 218.82 & 230.00 & 212.48 & 2364.77 \\
		F3 & 370 & 3.12 & 225.26 & 248.08 & 260.76 & 240.89 & 3039.51 \\
		F4 & 380 & 3.28 & 236.81 & 260.81 & 274.13 & 253.24 & 3359.24 \\
		F5 & 400 & 3.61 & 261.07 & 287.52 & 302.21 & 279.18 & 4082.72 \\
		F6 & 425 & 4.08 & 294.57 & 324.42 & 340.99 & 315.01 & 5197.74 \\
		F7 & 450 & 4.43 & 319.99 & 352.41 & 370.41 & 342.18 & 6133.28 \\
		F8 & 470 & 4.81 & 347.71 & 382.94 & 402.51 & 371.83 & 7242.13 \\  
		\bottomrule 
	\end{tabular} 
\end{table*}

\begin{figure}[tb]
	\centering
	%\vspace{-0.2cm}  %调整图片与上文的垂直距离
	%\setlength{\abovecaptionskip}{-0.05cm}   %调整图片标题与图距离
	%\setlength{\belowcaptionskip}{-2cm}   %调整图片标题与下文距离
	\includegraphics[width=0.49\textwidth]{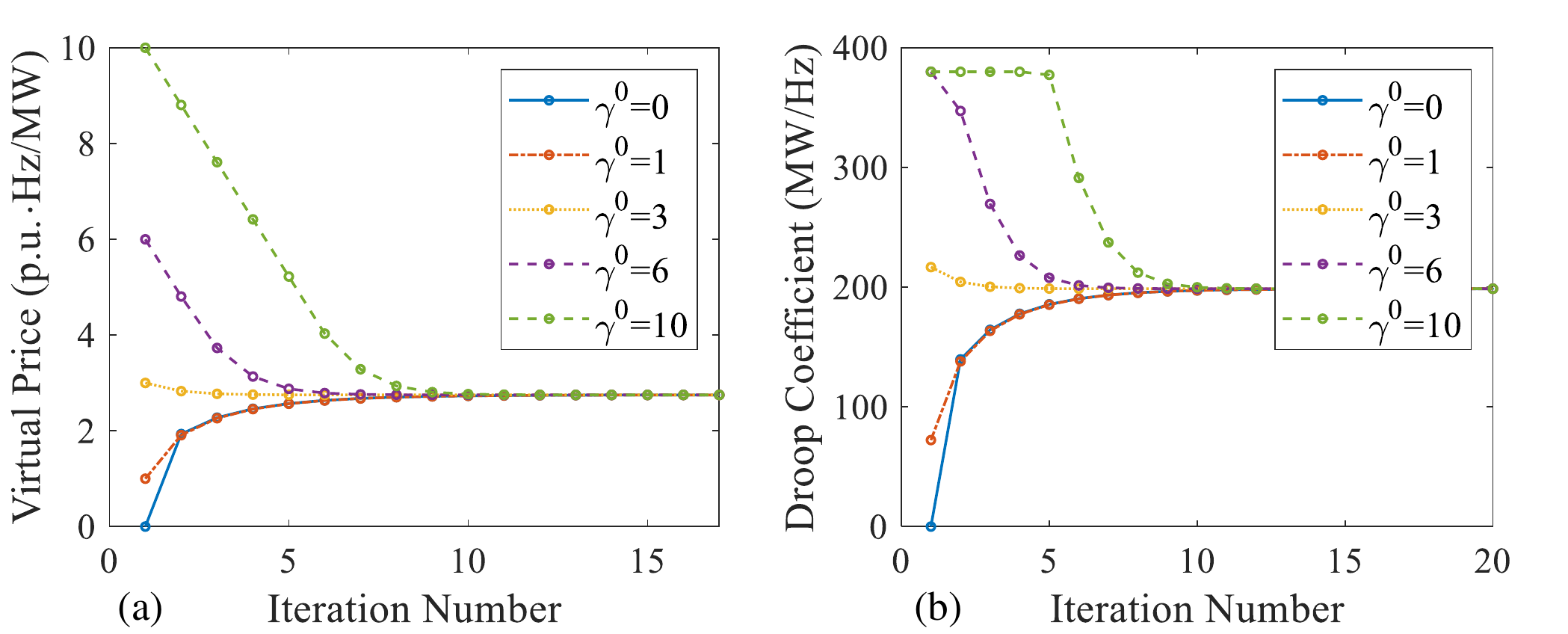}
	\caption{Iteration process. (a) virtual price. (b) droop coefficient of LCC1.}
	\label{case2}
\end{figure}

Moreover, to investigate the convergence of the SE-FP algorithm, we set five different initial values of the virtual price $\gamma^0$ between 0 p.u.$\cdot$Hz/MW to 10 p.u.$\cdot$Hz/MW to iteratively solve the Nash equilibrium with power imbalance $\Delta P_{E2}=$350 MW. The iteration processes of $\gamma$ and droop coefficient of LCC1 $k_1^D$ are shown in Fig. \ref{case2}. In Fig. \ref{case2}, with different initial value settings, the $\gamma$ and $k_1^D$ both converge to the same equilibrium, which verifies the convergence of the SE-FP algorithm. 

\subsection{Properties Verification}

\begin{figure}[tb]
	\centering
	%\vspace{-0.2cm}  %调整图片与上文的垂直距离
	%\setlength{\abovecaptionskip}{-0.05cm}   %调整图片标题与图距离
	%\setlength{\belowcaptionskip}{-2cm}   %调整图片标题与下文距离
	\includegraphics[width=0.42\textwidth]{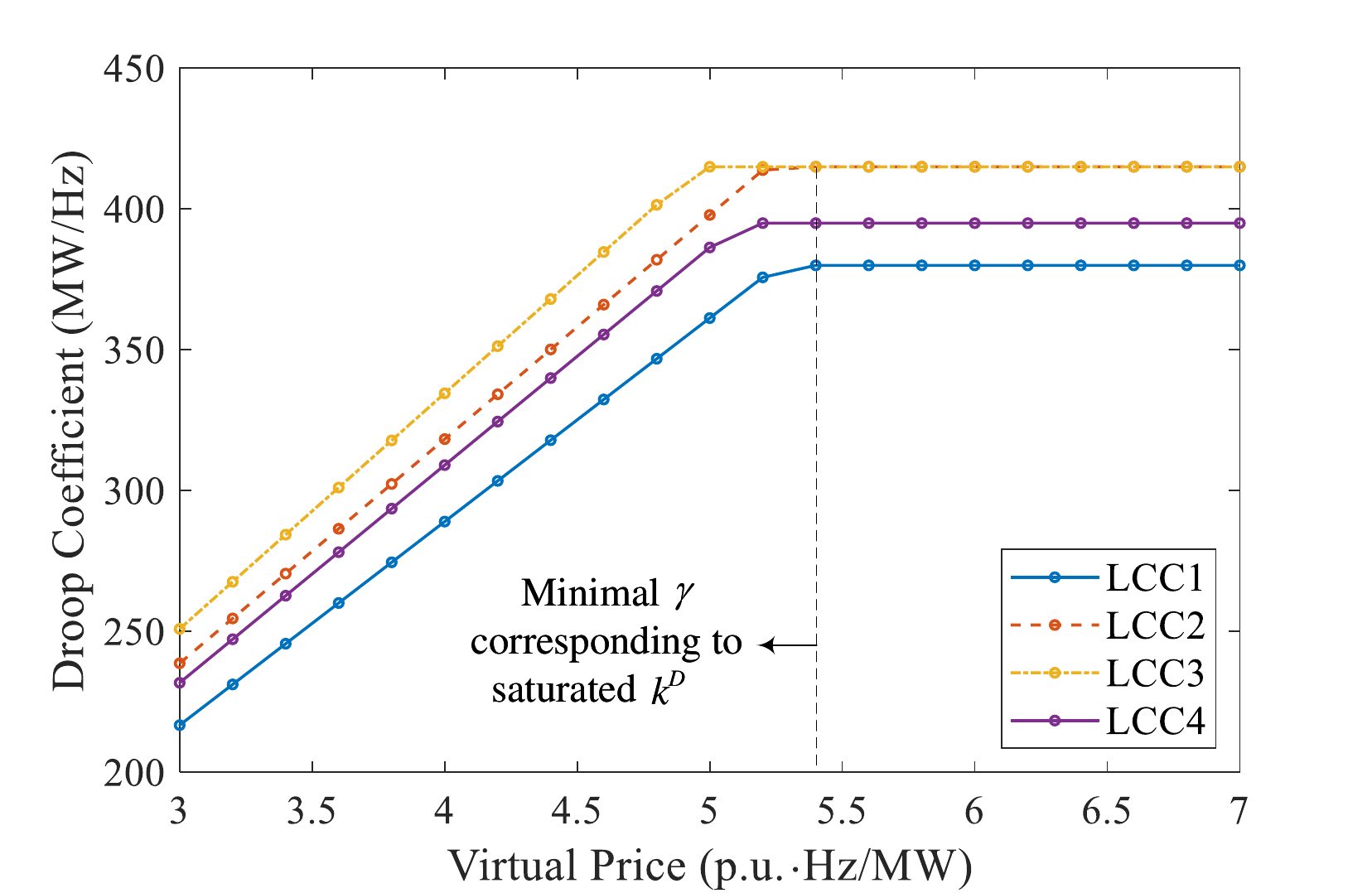}
	\caption{Optimal droop coefficients with different virtual price.}
	\label{case3}
\end{figure}

\begin{figure}[tb]
	\centering
	%\vspace{-0.2cm}  %调整图片与上文的垂直距离
	%\setlength{\abovecaptionskip}{-0.05cm}   %调整图片标题与图距离
	%\setlength{\belowcaptionskip}{-2cm}   %调整图片标题与下文距离
	\includegraphics[width=0.42\textwidth]{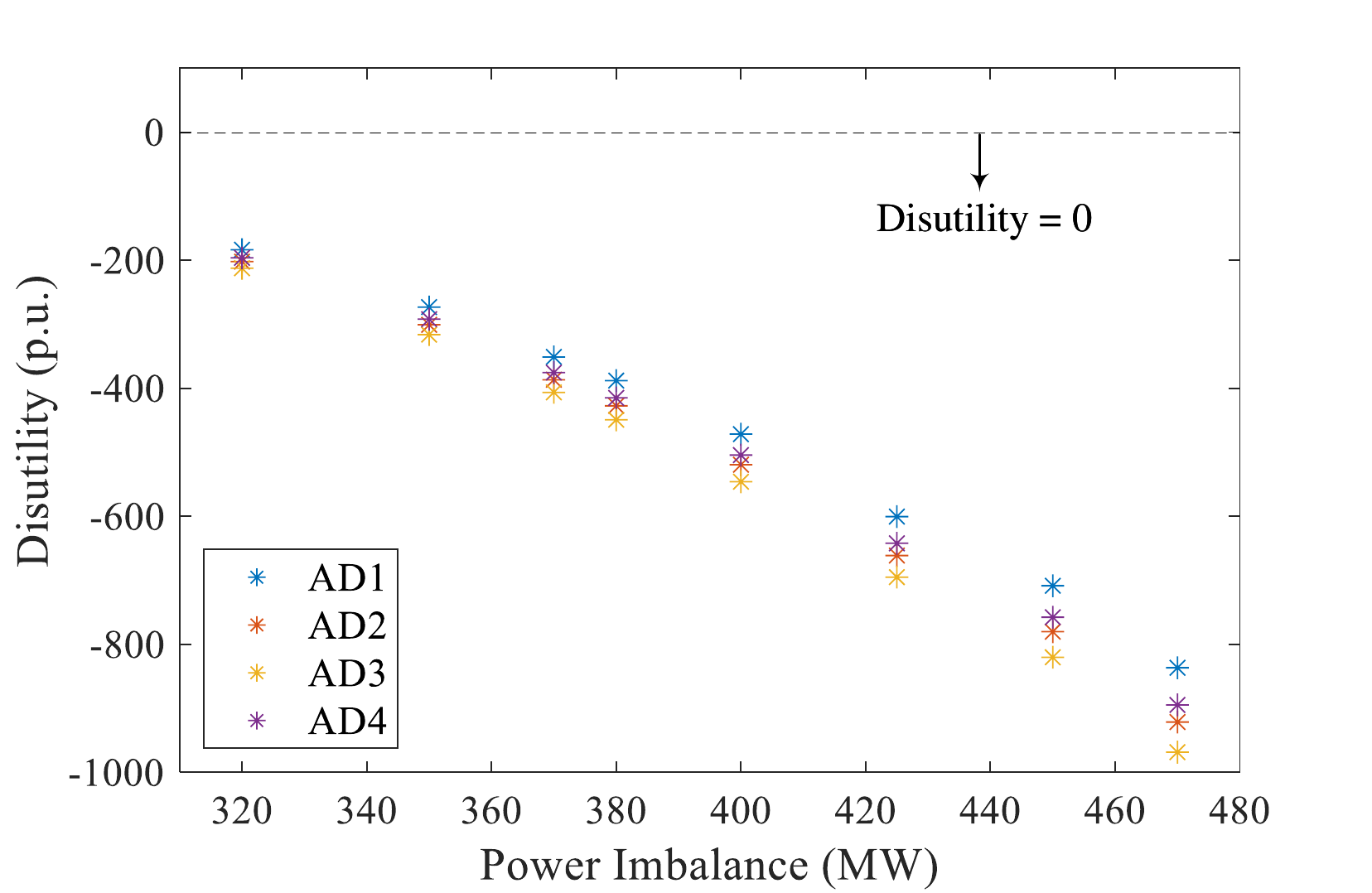}
	\caption{Disutilities of AD systems with variousc power imbalances.}
	\label{case4}
\end{figure}

We first verify the uniqueness of the Nash equilibrium by showing the monotonicity of the optimal droop coefficient with respect to the virtual price $\gamma$. We set $\gamma$ to vary from 3 p.u.$\cdot$Hz/MW to 7 p.u.$\cdot$Hz/MW, and the optimal droop coefficients of LCC-HVDC systems can be derived by solving the MOAD problems \eqref{moad_pro}, as shown in Fig. \ref{case3}. Fig. \ref{case3} not only shows the aforementioned monotonicity and uniqueness, but also illustrates that the optimal droop coefficients from MOAD problems tend to saturated values (upper bounds) as $\gamma$ continuously increases. Therefore, when all the droop coefficients reach their saturated values due to a considerable power imbalance, the incentive game can give the minimal reward $R$ corresponding to the minimal $\gamma$ in Fig. \ref{case3}, as mentioned in Remark 2.

Then, for each emergency fault in $\mathcal{F}$, we show the disutilities of the AD systems at the Nash equilibrium in Fig. \ref{case4} to verify the individual rationality of the proposed incentive mechanism. As shown in Fig. \ref{case4}, the disutilities of AD systems with various power imbalances are all less than zero, which means that by participating in the designed incentive mechanism, the AD systems can obtain more utilities than not. Thus, individual rationality is verified.  

\begin{figure}[tb]
	\centering
	%\vspace{-0.2cm}  %调整图片与上文的垂直距离
	%\setlength{\abovecaptionskip}{-0.05cm}   %调整图片标题与图距离
	%\setlength{\belowcaptionskip}{-2cm}   %调整图片标题与下文距离
	\includegraphics[width=0.49\textwidth]{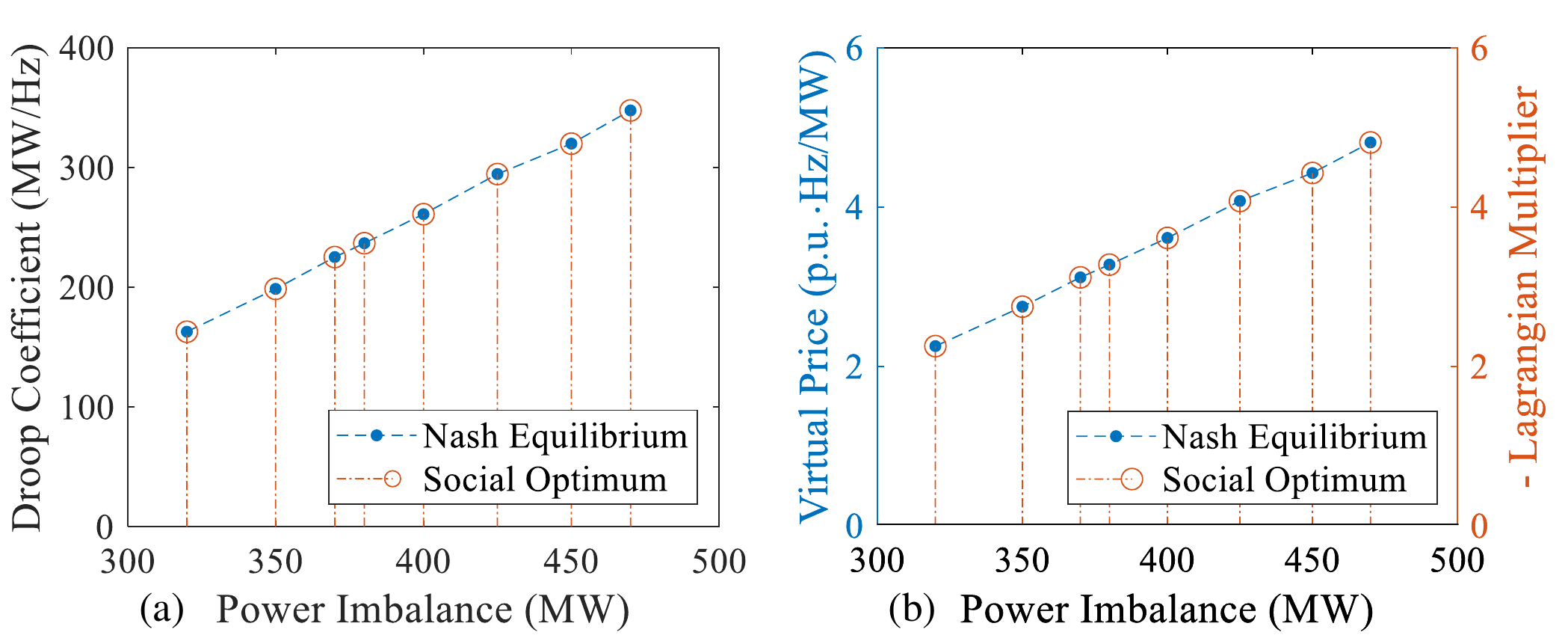}
	\caption{Comparison between Nash equilibrium and social optimum. (a) droop coefficient of LCC1. (b) virtual price and Lagrangian multiplier.}
	\label{case5}
\end{figure}

In addition, we compare Nash equilibriums and social optimums to verify the social optimality of the incentive mechanism, as shown in Fig. \ref{case5}, where the social optimums are derived by solving the social welfare problems \eqref{social_opt} with various power imbalances. In Fig. \ref{case5}(a), for each power imbalance, the LCC1 droop coefficient at Nash equilibrium is identical to that of the social optimum. And in Fig. \ref{case5}(b), the virtual price at Nash equilibrium is equal to the opposite of the Lagrangian multiplier of constraint \eqref{social_opt_2}. Thus, the social optimality of the proposed mechanism holds.  

\subsection{Impact of the Expected Frequency Deviation}

\begin{figure}[tb]
	\centering
	%\vspace{-0.2cm}  %调整图片与上文的垂直距离
	%\setlength{\abovecaptionskip}{-0.05cm}   %调整图片标题与图距离
	%\setlength{\belowcaptionskip}{-2cm}   %调整图片标题与下文距离
	\includegraphics[width=0.49\textwidth]{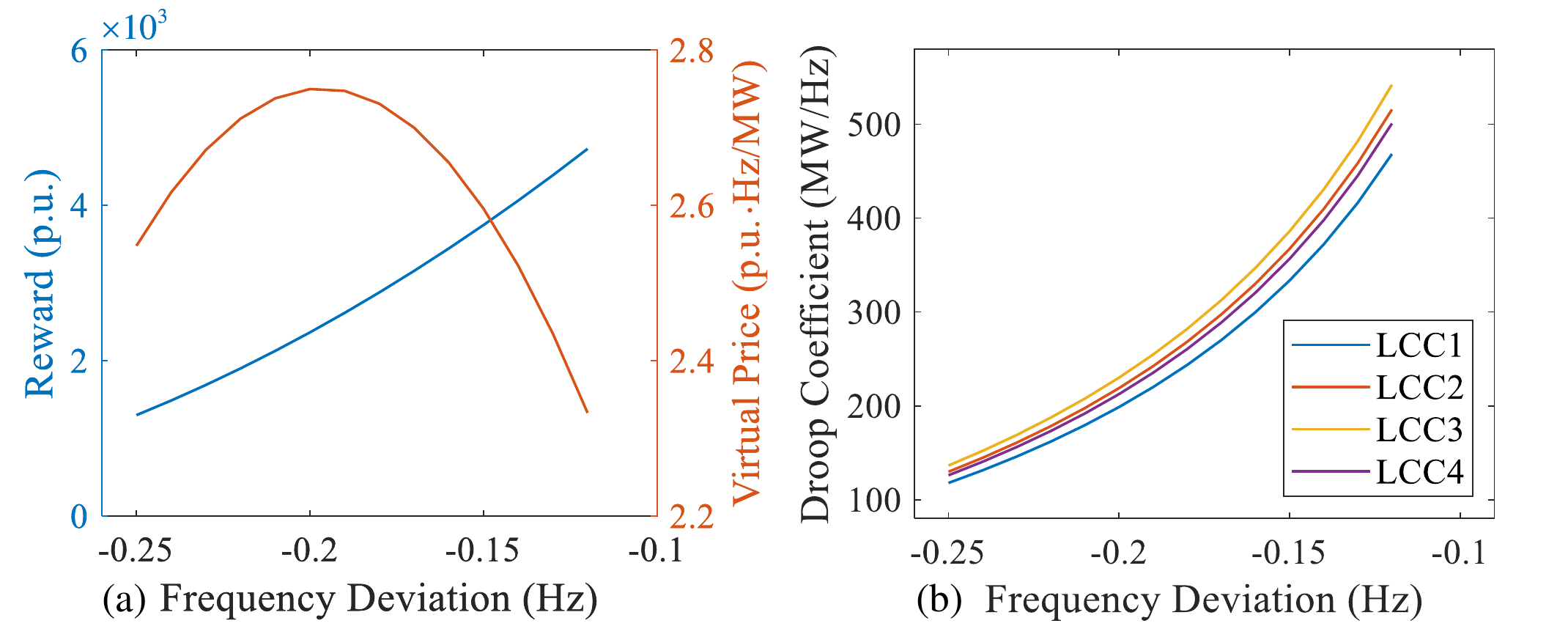}
	\caption{Nash equilibriums with various expected frequency deviations. (a) reward and virtual price. (b) droop coefficients of LCC-HVDC systems.}
	\label{case6}
\end{figure}

In Section III.B, to effectively seek the Nash equilibrium, we assume that the expected frequency deviation $\omega^{am}$ is a constant determined by the AM system. In this subsection, we further discuss the impact of $\omega^{am}$ on the Nash equilibrium of the incentive game, in order to provide reference for the AM system to determine the $\omega^{am}$.

We set $\omega^{am}$ to vary from $-$0.25 Hz to $-$0.12 Hz, and respectively solve the Nash equilibrium of the incentive game with power imbalance $\Delta P_{E2}=$350 MW. The related variables at Nash equilibrium are shown in Fig. \ref{case6}. As the $\omega^{am}$ increases, the reward $R$ in Fig. \ref{case6}(a) and the droop coefficients in Fig. \ref{case6}(b) also increase, while the virtual price $\gamma$ presents a tendency of increasing first and then decreasing. Referring to the above tendencies, the AM system can further determine the expected frequency deviation according to its own requirements.

\section{Conclusion}

In this paper, a control-parameter-based incentive mechanism is proposed to incentivize the LCC-HVDC systems and their connected adjacent AC systems to participate in the droop-based EFC of the MIDC system. Benefitting from the immediate adjustability of the droop coefficients of LCC-HVDC systems, the proposed incentive mechanism is able to deal with various possible emergency frequency faults. Then, to implement the proposed mechanism in the MIDC system, a non-cooperative-based incentive game model is formulated, and a fixed-point-based algorithm for seeking the Nash equilibrium is designed, which can be easily applied in engineering practice. The uniqueness of the Nash equilibrium is rigorously proven. And several preferable properties of the designed incentive mechanism, i.e., the individual rationality, incentive compatibility and social optimality, are analyzed and proven. In the case study for a MIDC test system, the effectiveness of the proposed incentive mechanism is illustrated, and the properties of the mechanism are verified.

\bibliographystyle{IEEEtran}
\bibliography{mybib}

\end{document}